\documentclass[twocolumn,showpacs,preprintnumbers,amsmath,amssymb]{revtex4}
\usepackage{graphicx}
\usepackage{dcolumn}
\usepackage{bm}
\usepackage{color}
\begin{document}

\title{Exploring origin of small x saturation in collinear approach }

\author{ A.M.~Snigirev$^{1,2}$ and G.M.~Zinovjev$^{3}$}

\affiliation{$^{1}$ Skobeltsyn Institute of Nuclear Physics, Lomonosov Moscow State University, 119991, Moscow, Russia }

\affiliation{$^{2}$ Bogoliubov Laboratory of Theoretical Physics, JINR, 141980, Dubna, Russia }

\affiliation{$^{3}$ Bogolyubov Institute for Theoretical Physics, National Academy of Sciences of Ukraine, Kiev 03143, Ukraine}

\date{\today}
\begin{abstract}
A modification of the collinear evolution equations as an appropriate approach to improve the behavior of parton distribution functions in the region of small longitudinal momentum fractions, and to find more theoretical arguments to clarify the possible appearence of saturation regime is suggested. It is argued that parton diffusion in the rapidity space at large parton densities along the space-time evolution could result in the emergence of a natural saturation scale on which freezing actually occurs.
\end{abstract}


\maketitle
\section{\label{sec1}Introduction}
Nowadays it is widely recognized the hadron interactions at very high energies are driven by the states with very high densities of partons (quarks and gluons), in particular, with small longitudinal momentum fractions $x$. The routine theoretical framework for analyzing such systems is essentially grounded on the QCD collinear factorization where the calculated cross sections are decomposed in the perturbative coefficient functions and nonperturbative parton densities of which evolvement is treated according the DGLAP equations~\cite{gribov, lipatov, dokshitzer, altarelli}. Already these linear equations capture qualitatively the traits associated with an increase in the gluon densities at small $x$ with extremely large $Q^2$ values. The latter turn out also quite instrumental, for example, to justify by neglecting any type of higher twist corrections and some perturbative resummation contributions. An idea to follow the evolution within the perturbative paradigm and to evaluate the leading contributions at small $x$ for not very large $Q^2$  led to the development of the BFKL approach associated to so-called high energy factorization. However, resolving the corresponding BFKL equations~\cite{bfkl, bfkl2, bal} exhibits a very strong raise (power-like) of the gluon density at small $x$ that is stronger than the experimental data analysis demonstrates and leads to apparent violation of unitarity at very small $x$. It signals some theoretical problems generated by appearance of an infrared instability related to a diffusion with the rapidity evolution and the consistent description of QCD coupling constant $\alpha_s$ behavior that should reflect a very sophisticated interplay of perturbative and nonperturbative QCD physics. Apparently, both look like an ensuing result of taking into account the linear evolution only with resummation in these approaches. From phenomenological point of view an observation  of a scaling law in wide range of small $x$ and $Q^2$ was done~\cite{praszalowicz} thereby demonstrating an onset of saturation scale. This fact is quite interesting because it may provide a perturbative scale in  high density region of small $x$ where linear evolution approximtion works and provides, in a sense, a boundary condition to the linear evolution equations. In fairness, remember it has been long time ago~\cite{ryskin, muelqui} argued that eventually the system under consideration should enter a new regime, where the rate of growing gluon density slows down and saturates possibly curing, thus, a potential conflict with unitarity of the underlying scattering. Actually, the restoration of the unitarity in high energy limit of QCD remains a challenging problem, although several approaches, drawing a scenario with nonlinear behavior, are being explored in past years (see, for example, 
\cite{lipatov97, mueller, mclerran, kovchegov, iancu, Gelis:2012ri, Dumitru:2012si, Blaizot:2016qgz} and references therein) but those allow us to conclude only that we are still no essentially closer to knowing where the problem solution lies besides of very general claim about the nonperturbative finite density effects which are left out entirely from the BFKL evolution. The interest in physics of high density regime of small $x$ QCD is greatly increasinng and dictated by an avalanche of experimental data on collisions of relativistic heavy ions overwhelming this area of research in the last decades.

Meanwhile, there is another opportunity to address the problem in the framework of well-known DGLAP approach that we would like to draw attention to in this letter.  It concerns one possible modificaion of collinear time-like evolution equations that has been also discussed long time ago~\cite{dremin1, dremin2} as well in the context of  increasing parton multiplicity in electron-positron annihilation into hadrons. We adapt this modification for the space-like evolution of parton distribution functions and demonstrate it develops a saturated regime of color glass condensate~\cite{Iancu}. This is a regime of strong color fields in which nonlinear dynamics come to the perceptible play and signals thereby an appearance of the natural saturation scale on which the evolution is, in fact, frozen, thus indicating also the universality of both phenomena. Actually, such an approach is treated as an effective theory of high energy scattering successefuly describing the data measured in experiments.

The paper is organized as follows. In Sec. II we review briefly the principal features of the DGLAP evolution necessary in further. The particular modified QCD evolution is discussed in Sec. III. In Sec. IV the extension to the double parton distribution functions is considered. The possible phenomenological issues are discussed in Sec. V, together with some conclusions.

\section{\label{sec2} Collinear evolution }

One may take the value of hard scale as the evolution variable in the DGLAP approach. The most popular choice is the transfer momentum squared $Q^2$, or its logarithm $\xi=\ln(Q^2/Q_0^2)$. The double logarithm that takes into account explicitly the behavior of the effective coupling constant in the leading logarithm approximation proves very instrumental as well 
\begin{eqnarray}
\label{t}
t = \frac{2}{\beta}\ln\Bigg
[\frac{\ln(\frac{Q^2}{\Lambda^2 })}
{\ln(\frac{Q_0^2}{\Lambda^2})}\Bigg],
\end{eqnarray}
where $\beta = (11N_c-2n_f)/3$ ${\rm {in~ QCD}}$, $Q_0$ is the some characteristic scale above which the perturbative theory is applicable, $n_f$ is the number of active flavors, $\Lambda$ is the QCD dimensional parameter and $N_c=3$ is the color number. In Eq.~(\ref{t}) the one loop running QCD coupling 
\begin{equation}
\label{alp}
\alpha_s(Q^2)=\frac{4\pi}{\beta\ln(Q^2/\Lambda^2)}
\end{equation}
was used. 

The DGLAP evolution equations~\cite{gribov,lipatov, dokshitzer, altarelli} assume the simplest form if we use the natural dimensionless evolution variable $t$; that is, 
\begin{equation}
\label{e1singl}
 \frac{dD_h^j(x,t)}{dt} = 
\sum\limits_{j{'}} \int \limits_x^1
\frac{dx{'}}{x{'}}D_h^{j{'}}(x{'},t)P_{j{'}\to j}\Bigg(\frac{x}{x{'}}\Bigg).
\end{equation}
These equations describe the evolution of single distributions $ D^j_h(x,t)$ of bare quarks, antiquarks and gluons ($j = q, {\bar q}, g$) within a hadron $h$ in response to the change of evolution variable $t$. The kernels, $P$, of these equations include a regularization at $x \rightarrow x{'}$ and are known in their appropriate forms.

Equations~(\ref{e1singl}) are explicitly solved by introducing the Mellin transforms
\begin{eqnarray}
\label{mellin}
  M_h^{j}(n,t) = \int\limits_{0}^{1}dx x^n~D_{h}^{j}(x,t),
\end{eqnarray}
which reduce those to a system of ordinary linear-differential equations at the first order:
\begin{equation}
\label{1.19}
dM_h^j (n,t)/dt = \sum\limits_{j{'}} M_h^{j{'}}(n,t) P_{j{'} \to j}(n), 
\end{equation}
where
\begin{eqnarray}
\label{1.21}
P_{j{'} \to j}(n)  =  \int\limits_0^1 x^n P_{j{'} \to j} (x)dx.   
\end{eqnarray}

In order to obtain the distributions in $x$ representation the inverse Mellin transformation should be performed
\begin{eqnarray}
\label{mellin in}
x D_h^{j}(x,t) =~\int\frac {dn}{2\pi i} x^{-n}~M_{h}^{j}(n,t),
\end{eqnarray}
where the integration runs along the imaginary axis to the right from all $n$ singularities. It can be done  in a general form numerically only. However, the asymptotic behavior can be estimated in some interesting and simple enough limits with the technique under consideration.

The solutions of the DGLAP equations with the given initial conditions $D^j_h (x,0)$ at the reference scale $Q_0 (t=0)$  can be expressed by the Green functions $D^j_{i}(z,t)$ in the following way:
\begin{eqnarray}
\label{1solution}
 D_h^j(x, t)
 = \sum\limits_{i{'}} \int \limits_x^1
\frac{dz}{z}~D_h^{i}(z,0)~D_{i}^j(\frac{x}{z},t).
\end{eqnarray}
These Green functions (gluon distributions at the parton level) $D^j_{i}(z,t)$ are the solutions of Eqs.~(\ref{e1singl}) at the parton level with the singular initial conditions $D^j_{i}(z,t=0)= \delta (x-1)\delta_{ij}$ and in the double logarithm approximation (see, for instance, \cite{dokshitzer,ryskin}) look like
\begin{eqnarray}
\label{mellin in x=0}
 x D_g^{g}(x,t) & &= 4N_c t\exp{[-at]} I_1(v)/v \nonumber \\
& & \simeq 4N_ct v^{-3/2}\exp{[v-at]}/\sqrt{2\pi},
\end{eqnarray}
where
\begin{eqnarray}
v=\sqrt{8N_ct\ln{(1/x)}}, a=\frac{11}{6}N_c +\frac{1}{3}n_f/N^2_c,
\end{eqnarray}
and $I_1$ is the standard modified Bessel function. This result just illustrates the unitarity violation at very small $x$. In addition, one should also note that the mean number of partons of type $j$ in a parton of type $i$
\begin{eqnarray}
\label{solution n}
< n^j >_i = M_i^{j}(0,t) = [\exp{P(0) t}]_i^j
\end{eqnarray}
can not be correctly determined in the collinear approach because the kernels $P_{g \to g}(0)$ and $P_{q\to g}(0)$ are divergent and some improvements are necessary to be done at very small $x$.

\section{\label{sec3} Collinear evolution with dissipation}

The modification of collinear time-like evolution equations was discussed in Refs.~\cite{dremin1, dremin2} to take into account the formation (so-called pionization) of soft quark-antiquark pairs at a hard quark (gluon) propagating. In analogy with the electron-photon showers the energy outflow was phenomenologically simulated by the dissipative terms in the evolution equations with rather interesting income. Such a modification for the space-like evolution has, of course, another physical motivation in our case due to the parton diffusion in the rapidity space at large parton densities, and the following evolution equations are suggested:
\begin{eqnarray}
\label{e1singl-d}
 \frac{\partial D_i^j(x,t)}{\partial t} &=& 
\sum\limits_{j{'}} \int \limits_x^1
\frac{dx{'}}{x{'}}D_i^{j{'}}(x{'},t)P_{j{'}\to j}\Bigg(\frac{x}{x{'}}\Bigg) \nonumber \\
& & + \gamma^j\frac{\partial D_i^j(x,t)}{\partial x}
\end{eqnarray}
with $\gamma^j$ as some parameters characterizing the process of the energy outflow. 

In the situation of small dissipation, $\gamma^j \ll 1$, the mean number of partons can be calculated~\cite{dremin2} by using the Mellin technique. For credibility we bring here the result for gluon multiplicity at the early evolution stage ($t \ll 1$) only referring to the transparent, but laborious, calculations performed in~\cite{dremin2}:
\begin{eqnarray}
\label{solution g}
 &  &< n^g >_g = I_0(V)e^{-at} \nonumber \\
& & + \sqrt{\frac{2N_c t}{\ln{(1/\gamma^g)}}}\ln{\sqrt{\frac{\ln{(1/\gamma^g)}}{2N_ct}}} I_1(V)e^{-at},
\end{eqnarray}
where
\begin{eqnarray}
V=\sqrt{8N_ct\ln{(1/\gamma^g)}},  
\end{eqnarray}
and $I_{0}$ is another modified Bessel function. This result (\ref{solution g}) reproduces exactly the mean number of gluons with the longitudinal momentum fractions larger than $x_0 = \gamma^g$ as calculated in the DGLAP unmodified approach. The exercise above makes transparent the physical meaning of the dissipative term. It establishes the scale of energy drift because gluons (partons) with the longitudinal momentum fractions less than $ \gamma^g$ are simply withdrawn from consideration. Moreover, the evolution is, in fact, frozen at the scale~\cite{dremin1}
\begin{eqnarray}
Q^2_{fr}= \Lambda^2 (Q^2/\Lambda^2)^{\gamma^g}.  
\end{eqnarray}
The origin of this freezing scale is similar to the saturation scale in the color glass condensate (CGC) approach~\cite{mclerran, iancu, Gelis:2012ri, Dumitru:2012si, Blaizot:2016qgz, Iancu}.

\section{\label{sec4} Generalizing to double parton distributions}

The extension of basic equations to double parton distribution functions is straightforward:
\begin{eqnarray}
\label{edouble}
& &\frac{\partial D_h^{j_1j_2}(x_1,x_2,t)}{\partial t} \\
& &=\sum\limits_{j_1{'}} 
\int\limits_{x_1}^{1-x_2}\frac{dx_1{'}}{x_1{'}}D_h^{j_1{'}j_2}(x_1{'},x_2,t)
P_{j_1{'}
\to j_1} \Bigg(\frac{x_1}{x_1{'}}\Bigg) \nonumber\\
& & + \gamma^{j_1}\frac{\partial D_h^{j_1j_2}(x_1,x_2,t)}{\partial x_1}  \nonumber\\
& &+ \sum\limits_{j_2{'}}\int\limits_{x_2}^{1-x_1}
\frac{dx_2{'}}{x_2{'}}D_h^{j_1j_2{'}}(x_1,x_2{'},t)P_{j_2{'} \to j_2}
\Bigg(\frac{x_2}{x_2{'}}\Bigg) \nonumber\\
& & + \gamma^{j_2}\frac{\partial D_h^{j_1j_2}(x_1,x_2,t)}{\partial x_2}  \nonumber\\
& &+ \sum\limits_{j{'}}D_h^{j{'}}(x_1+x_2,t) \frac{1}{x_1+x_2}P_{j{'} \to
j_1j_2}\Bigg(\frac{x_1}{x_1+x_2}\Bigg).\nonumber
\end{eqnarray}
Here, the splitting kernels,
\begin{equation}
\frac{1}{x_1+x_2} P_{j{'} \to
j_1j_2}(\frac{x_1}{x_1+x_2}),
\end{equation}
which appear in the nonhomogeneous part of the equations, are the nonregularized one-loop well-known DGLAP kernels without the ``+'' prescription.  
The unmodified equations were derived first in Refs.~\cite{Kirschner:1979im,Shelest:1982dg} in framework of the DGLAP approach. The functions $D_h^{j_1j_2}(x_1,x_2,t)$ in question have a specific interpretation in the leading logarithm approximation of perturbative QCD. They are the inclusive probabilities which allow one to find two bare partons of types $j_1$ and $j_2$ with the given longitudinal momentum fractions $x_1$ and $x_2$ in a hadron $h$. 

The dissipative terms provide the energy outflow and establish the scale of energy drift as well. Gluons (partons) with the longitudinal momentum fractions less than $ \gamma^j$ are simply removed again from consideration for each of two parton cascade branches practically independently. In the small $x$ region we can restrict ourselves to homogeneous evolution equations because the solutions of nonhomogeneous unmodified equation are substantial at not parametrically small longitudinal momentum fractions only~\cite{Snigirev:2014eua}. Moreover, the homogeneous evolution equations (independent evolution of two branches) admit the factorization of double parton distribution functions:
\begin{eqnarray} 
\label{DxD_Q}
 D^{j_1 j_2}_{h}(x_1, x_2,t)
 \simeq D^{j_1}_h (x_1,t) D^{j_2}_h (x_2,t)
\end{eqnarray}
as a good approximate solution, if such a factorization was assumed at the reference scale  $ Q_0 (t=0)$.

Further we hold the leading exponential terms only if those have the same structure~\cite{Ryskin:2012qx} both at the parton level and the hadron level under smooth enough initial conditions at the reference scale. Indeed, Eq. (\ref{1solution}) in the double logarithm approximation reads 
\begin{eqnarray}
\label{1gsolution}
 x D_h^g(x; t) & & \simeq
\int \limits_0^Y dy[zD_h^{g}(z,0)]|_{1/z=\exp{y}}\nonumber\\
& & \times \exp{[\sqrt{8N_c}\sqrt{t(Y-y)} ]}\nonumber\\
 & & \sim \exp{[\sqrt{8N_c}\sqrt{tY} ]}
\end{eqnarray}
with  $Y=\ln(1/x)$.
The $y$ integration is not as a saddle-point type, and, therefore, one of the edges, just $y\to 0$ ($z\to 1$), dominates, provided that the initial gluon distribution does not increase too much with $z$ decreasing. Actually, one needs  $zD_h^{g}(z,0) \sim (1/z)^a$ at $z \to 0$ with $a<A$, where $A=\sqrt{2N_ct/Y} >0$. Let's notice that the parametrization of the initial gluon distributions, usually used, satisfies this condition (e.g., the CTEQ parametrization of Ref. ~\cite{cteq}). Thus, as a result we have for the double gluon distributions~\cite{Ryskin:2012qx} in this appproximation:
\begin{eqnarray}
\label{saddle1x2} 
& & x_1x_2 D^{gg}_{h}(x_1,x_2,t)  \nonumber \\
& & \sim
\exp{[\sqrt{8N_c} (\sqrt{t\ln{(1/x_1)}}+\sqrt{t\ln{(1/x_2)}})]}
\end{eqnarray}
with the infinite mean number of such gluons. If the two branches evolves independently then by introducing the dissipative terms slow down the rate of gluon density increase and one gets the {\it finite} mean gluon numbers as
\begin{eqnarray}
\label{n_gg} 
<n^{gg}>_h \sim
\exp{[\sqrt{8N_c} (\sqrt{t \ln{(1/\gamma^g)}}+\sqrt{t\ln{(1/\gamma^g)}})]}
\end{eqnarray}
since the gluons with the longitudinal momentum fractions less than $ \gamma^g$ are simply excluded.

\section{\label{sec5}Discussion and summary}

Clearly, the dissipative parameters above can not be determined within the DGLAP approach. They are treated as the phenomenological parameters in numerical simulations and shoud be estimated in the other models for further applications. The phenomena of saturation and slowing down an increase of gluon density take place also in the CGC scenario~\cite{mclerran, iancu, Gelis:2012ri, Dumitru:2012si, Blaizot:2016qgz}. However, the saturation scale is energy dependent in that approach and, nevertheless, it comes about quite predictive. For example, in the Golec-Biernat-Wusthoff (GBW) model~\cite{gbw, gbw1} it is parametrized by three parameters :
\begin{eqnarray}
\label{qs}
Q^2_{s}= Q_0^2 (x_0/x)^{\lambda},
\end{eqnarray}
with $Q_0=1$ GeV, $x_0\simeq 0.0001$, and $\lambda \simeq 0.3$ which have been used to describe accurately the HERA data~\cite{hera}. The value of characteristic energy (longitudinal momentum fraction) $x_0$ in Eq.~(\ref{qs}) allows us to estimate the dissipative parameter $\gamma^g$ that has a physical meaning similar to $x_0$. In fact, it justifies the assumption of small dissipation used in the previous Sections to obtain the crucial estimates~(\ref{solution g}) and (\ref{n_gg}) which are pretty encouraging to investigate the properties of modified collinear equations further as 
a new alternative insight into the saturation physics extending the initial limits of linear approach.


In summary, the modified collinear evolution equations are suggested to extract information on the properties of hot and dense QCD medium created in the experiments on heavy ion collisions searching the quark-gluon plasma, a thermolized phase, that may exist in very spesific regimes for very short periods of time. Comprehensive phenomenological analysis of proton-proton collisions based on the QCD factorization, as a key instrument, made it possible to extract the universal distribution functions validating such an approximation and open up (quite often) transparent ways for introducing the efficient corrections. Truly, these corrections at leading power of the large momentum transfer are fairly general and easily traceable but the corrections within the factorized forms turn out very complicated and too much sensitive to the process details, as it was shown again many years ago, because of the QCD multiple scaterrings which differ~\cite{zakharov} hadronic and heavy ion collisions significantly. The model presented in this letter shows a possibility of forming perturbatively a dynamical regime in particular kinematical configuration which could not be foreseen according to the theoretical dogmas. As argued, it concerns a regime of high parton densities and dynamical interactions described definitely by nonlinear equations. The evolution of hadron scattering amplitudes, at least, in the framework of color dipole picture in such a regime is quite similar~\cite{peschanski} to the time evolution of a classical particles system undergoing reaction-diffusion processes. Amazingly, by introducing the dissipative terms results in an origin of natural saturation scale on which the evolution is frozen and the gluons with longitudinal momentum less than $\gamma^g$ are simply excluded. In the phenomenological applications the direct numerical solutions of suggested modified equations may occur simpler than the BFKL treatment of very small $x$ region.

\begin{acknowledgments}
Inspiring discussions (many years ago) with I.M.~Dremin, A. Kovner, A.V.~Leonidov and L.~McLerran are gratefully acknowledged.
The work of G.M. is supported by the Goal-Oriented Program of Cooperation between CERN and National Academy of Science of Ukraine ``Nuclear Matter under 
Extreme Conditions'' (agreement CC/1-2019, No.0118U005343). 
\end{acknowledgments}

\end{document}